\documentclass[twocolumn,prb,aps,showpacs,superscriptaddress]{revtex4-1}%
\usepackage{amsmath}
\usepackage{amssymb}
\usepackage{amsfonts}
\usepackage{bm}
\usepackage{amssymb}
\usepackage[dvips]{graphicx}
\usepackage{dcolumn}
\usepackage{txfonts}
\usepackage{makeidx}
\usepackage{color}
\usepackage{mathtools}
\usepackage{threeparttable}
\usepackage[linkcolor=blue,anchorcolor=black,citecolor=blue]{hyperref}

\begin{document}
\title{Electron Spin Decoherence in Silicon Carbide Nuclear Spin Bath}

\author{Li-Ping Yang }
\affiliation{Beijing Computational Science Research Center, Beijing 100084, China}

\author{Christian Burk}
\affiliation{3. Physikalisches Institut and Research Center SCOPE, University of Stuttgart, Pfaffenwaldring 57, 70569 Stuttgart, Germany}

\author{Mattias Widmann}
\affiliation{3. Physikalisches Institut and Research Center SCOPE, University of Stuttgart, Pfaffenwaldring 57, 70569 Stuttgart, Germany}

\author{Sang-Yun Lee}
\affiliation{3. Physikalisches Institut and Research Center SCOPE, University of Stuttgart, Pfaffenwaldring 57, 70569 Stuttgart, Germany}

\author{J\"{o}rg Wrachtrup}
\affiliation{3. Physikalisches Institut and Research Center SCOPE, University of Stuttgart, Pfaffenwaldring 57, 70569 Stuttgart, Germany}

\author{Nan Zhao}
\email{nzhao@csrc.ac.cn}
\homepage{http://www.csrc.ac.cn/~nzhao/}
\affiliation{Beijing Computational Science Research Center, Beijing 100084, China}
\affiliation{3. Physikalisches Institut and Research Center SCOPE, University of Stuttgart, Pfaffenwaldring 57, 70569 Stuttgart, Germany}
\affiliation{Synergetic Innovation Center of Quantum Information and Quantum Physics,
University of Science and Technology of China, Hefei 230026, China}


\begin{abstract}
In this paper, we study the electron spin decoherence of single defects in silicon carbide (SiC)
nuclear spin bath. We find that, although the natural abundance of $^{29}\rm{Si}$
($p_{\rm{Si}}=4.7\%$)
is about 4 times larger than that of $^{13}{\rm C}$ ($p_{\rm{C}}=1.1\%$),
the electron spin coherence time of defect centers in SiC nuclear spin bath in strong magnetic
field ($B>300~\rm{Gauss}$) is longer than that of nitrogen-vacancy (NV) centers
in $^{13}{\rm C}$ nuclear spin bath in diamond. The reason for this counter-intuitive result is
the suppression of heteronuclear-spin flip-flop process in finite magnetic field. Our results
show that electron spin of defect centers in SiC are excellent candidates for solid state
spin qubit in quantum information processing.
\end{abstract}
\maketitle

\section{introduction}

Investigations of nitrogen-vacancy (NV) centers in diamond have achieved significant progress in
the past years. One of the most promising properties of NV center is the long spin
coherence time~\cite{Science1997_Wrachtrup,NatNano2001_Wrachrup,JPCM06_Wrachtrup,PRL06_Lukin,Nature11_NZhao,PRB12_NZhao}
($\sim 10^2~\rm{\mu s}$ in samples with natural abundance nuclei, and even longer in isotope purified samples) even at room temperature. 
The  electron spin decoherence in pure diamond samples (e.g., type-IIb diamond) is caused by the magnetic fluctuations from
$^{13}\rm{C}$ nuclear spins (with natural abundance $p_{\rm{C}}=1.1\%$). Recently, experiments to
explore similar defect centers in different host materials have been started. Several types of related defect
 centers~\cite{PRL06_Son,PRB09_Umeda,PNAS10_Weber,PRB11_Baranov,PRL12_Soltamov,NatPhys14_Kraus,arxiv14_Wrachtrup}
in SiC have attracted great interest due to their outstanding features, such as weak
spin-orbit coupling~\cite{NatMat14_Ohshima}, wide band-gap~\cite{PRB00_Sorman,Nature_ Awschalom},
high thermal conductivity, and mature fabrication techniques, etc. At the same time,
some of these defect centers have shown to have non-zero spin for the orbital ground state\cite{Nature_ Awschalom},
which can be used as spin qubits. 
The host material SiC contains both $^{13}\rm{C}$ and $^{29}\rm{Si}$ nuclear spins, while $^{29}\rm{Si}$ nuclei have about 4 times larger natural abundance ($p_{\rm{Si}} = 4.7\%$) than $^{13}\rm{C}$ nuclei, which may imply faster spin decoherence.
However, very recent experiments show that the single defect center spin in SiC can evolve coherently for at 
least~$160~{\rm \mu s}$~\cite{arxiv14_Wrachtrup}, and the ensemble averaged coherence time could reach more than $1~{\rm ms}$~\cite{arxiv14_Awschalom} 
at cryogentic temperature. With all these progresses, for both quantum information applications and decoherence physics, 
a systematic study of coherence time and decoherenece mechanisms of defects in SiC is highly desirable.

In this paper, we demonstrate a counter-intuitive result, namely that the electron spin coherence time in SiC nuclear
spin bath is longer than that of NV centers in diamond in high magnetic field.
In particular, taking the silicon vacancy defects (denoted as $\rm{V}_{\rm{Si}}$ ) in 4H-SiC as an example, we perform microscopic calculations of the electron spin coherence time, and analyze the underlying decoherence processes.
As well-studied in various similar systems,
central spin decoherence in strong magnetic fields is mainly caused by  flip-flop of nuclear
spin pairs.
We show that, being different from the homonuclear spin pair cases ($^{13}{\rm C}$-$^{13}{\rm C}$
and $^{29}{\rm Si}$-$^{29}{\rm Si}$ pairs), the heteronuclear spin pair flip-flop (i.e., $^{29}{\rm Si}$-$^{13}{\rm C}$)
is significantly suppressed in strong fields, which is the key point for understanding the longer
$T_{2}$ time of defect centers in SiC nuclear spin bath.

The paper is organized as follows. Section~\ref{sec:Model} gives
the microscopic model of defect centers in 4H-SiC. The
numerical results and discussion are presented in Sec.~\ref{sec:results}.
Section~\ref{sec:conclusion} gives the conclusion.

\begin{figure}[b]
\includegraphics[width=8cm]{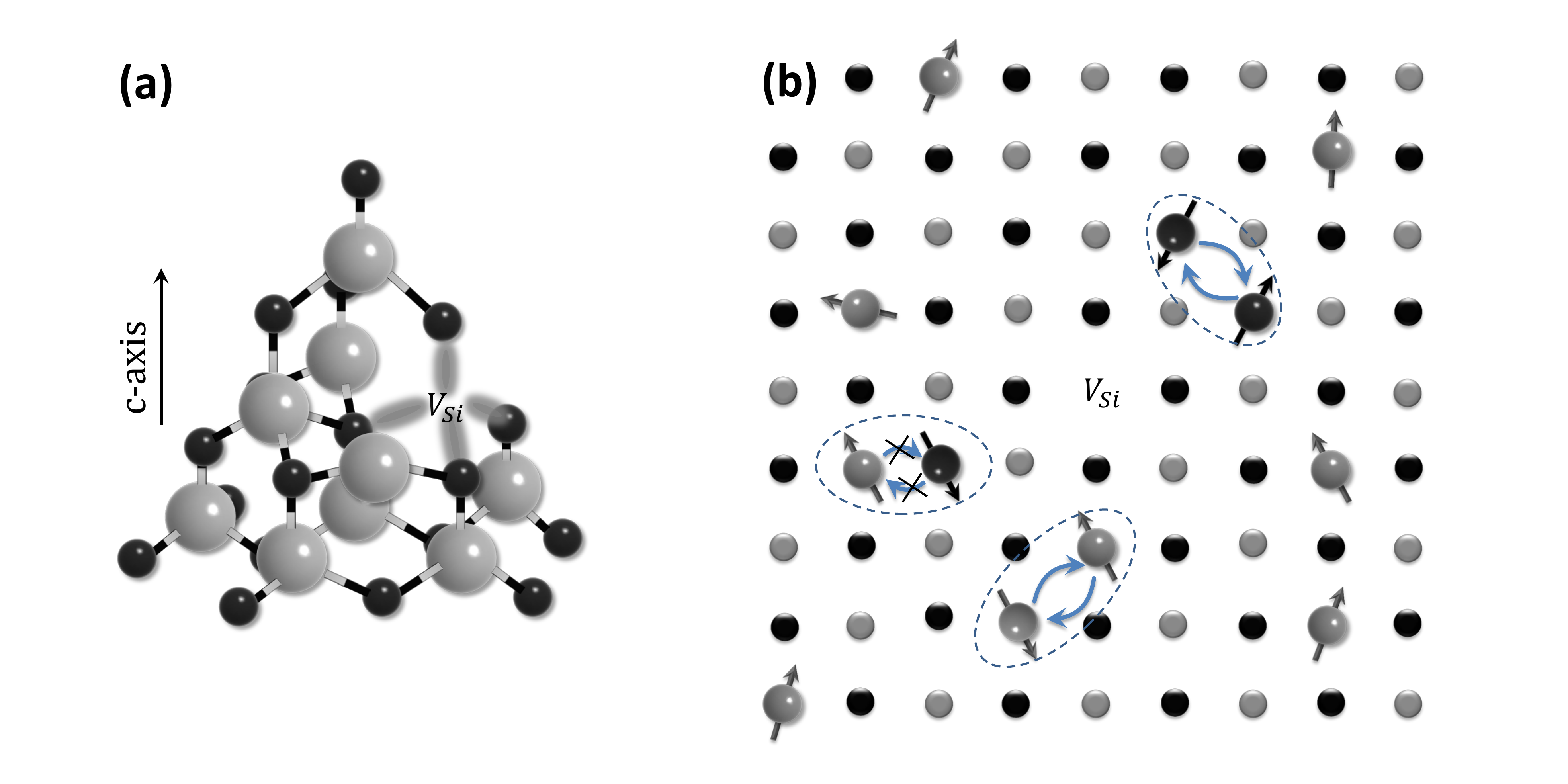}

\caption{\label{fig:schematic}(Color online) (a) Schematic for the
${\rm V_{Si}}$ defect in 4H-SiC. The $z$-direction is chosen along the $c$-axis of the crystal. Black and Gray spheres represent carbon and silicon atoms respectively.
(b) Schematic for SiC nuclear spin bath. The flip-flop process between the heteronuclear spins is significantly suppressed in finite magnetic fields.}
\end{figure}

\section{Model and heteronuclear spin pair dynamics\label{sec:Model} }
\subsection{Microscopic Model}
Recent experiments show various defect centers in SiC, such as ${\rm V_{Si}}$ and divacancies (denoted as ${\rm V_{Si}}$-${\rm V_{C}}$).
Here, we focus on ${\rm V_{Si}}$ in 4H-SiC [see Fig.~\ref{fig:schematic} (a)], where
coherent manipulation of single defect centers is achieved experimentally~\cite{arxiv14_Wrachtrup}.
The coherence time does not change significantly for different types of defect centers,
particularly in strong magnetic field (e.g., $B>300~\text{Gauss}$). The orbital ground state of
${\rm V_{Si}}$ defect centers in 4H-SiC is a quartet state with
$S=3/2$. There are two kind of ${\rm V_{Si}}$ centers in 4H-SiC, $\rm{T_{V1}}$ and
$\rm{T_{V2}}$, corresponding to two inequivalent lattice cites~\cite{PRB00_Sorman}. In this paper,
we take the $\rm{T_{V2}}$ center as an example to demonstrate physical
mechanism of the long decoherence time for vacancy centers in SiC.

Two types of nuclear spins, ${\rm ^{13}C}$ and ${\rm ^{29}Si}$ with
natural abundance $p_{\rm{C}}=1.1\%$ and $p_{\rm{Si}}=4.7\%$ respectively,
contribute to the decoherence of $\rm{T_{V2}}$ centers.
The central electron spin decoherence in an applied magnetic field $\mathbf{B}$
is caused by the magnetic fluctuations from a large number of nuclear spins
[see Fig.~\ref{fig:schematic} (b)], which is described by the following Hamiltonian
\begin{equation}
H=H_{\rm{es}}+H_{{\rm bath}}+H_{{\rm int}},
\end{equation}
where the electron spin Hamiltonian $H_{\rm{es}}$ is
\begin{equation}
H_{\rm{es}}=-\gamma_{e}\mathbf{B}\cdot \mathbf{S}+D S_{z}^{2}
\end{equation}
with electron gyromagnetic ratio $\gamma_{e}=-1.76\times10^{11}~{\rm rad\cdot s^{-1}\cdot T^{-1}}$,
and the zero-field splitting (ZFS) of $\rm{T}_{\rm{V}2}$ center denoted as $D$.
Recent experiments show that the ZFS  is in the range of  $10~{\rm MHz}- 100~{\rm MHz}$~\cite{PRB00_Sorman,NatPhys14_Kraus,arxiv14_Wrachtrup,NatMat14_Ohshima}.
Although the ZFS of $\rm T_{V2}$ is much smaller than that of NV center spin in diamond ($2.87{\rm GHz}$), it is still large enough to prevent the electron spin flipping due
to the weak hyperfine coupling (the typical hyperfine coupling strength $\lesssim 10^2~\rm{kHz}$).
Here, we take $D=35{\rm MHz}$~\cite{arxiv14_Wrachtrup} and assume the magnetic field direction is along
the $c$-axis of 4H-SiC as shown in Fig.~\ref{fig:schematic} (a), which is defined as the $z$-direction.

The Hamiltonian of the bath nuclear spins is
\begin{equation}
H_{{\rm bath}}=-\sum_{(i,\xi)}\gamma_{\xi}BI_{z}^{(i,\xi)}+H_{{\rm dip}},
\end{equation}
where the composite index $(i,\xi)$ denotes the $i$th nuclear spin of type $\xi$
with $\xi\in\{{\rm C,Si}\}$ and $i=1,2,\dots,N_{\xi}$ ($N_{\xi}$ is the number
of the $\xi$-type nuclear spin), and $\gamma_{\rm C}=6.73\times10^{7}~{\rm rad\cdot s^{-1}\cdot T^{-1}}$
and $\gamma_{\rm Si}=-5.32\times10^{7}~{\rm rad\cdot s^{-1}\cdot T^{-1}}$ are
the gyromagnetic ratios of ${\rm ^{13}C}$ and ${\rm ^{29}Si}$ nuclear
spins, respectively. Nuclear spins are coupled by magnetic dipole-dipole interaction
of the form
\begin{equation}
H_{{\rm dip}}=\frac{1}{2}\sum_{(i,\xi)\neq(j,\xi')}\mathbf{I}^{(i,\xi)}\cdot\mathbb{D}_{i\xi,j\xi'}\cdot\mathbf{I}^{(j,\xi')},
\end{equation}
where
\begin{eqnarray*}
\mathbb{D}_{i\xi,j\xi'} & = & \frac{\mu_{0}\gamma_{\xi}\gamma_{\xi'}}{4\pi r_{ij}^{3}}\left(1-\frac{3 \mathbf{r}_{ij} \mathbf{r}_{ij} }{r_{ij}^{2}}\right)
\end{eqnarray*}
is the dipolar coupling tensor between two nuclei located at $\mathbf{r}_{(j,\xi)}$ and $\mathbf{r}_{(i,\xi')}$.
The relative displacement between them is  $\mathbf{r}_{ij}=\mathbf{r}_{(j,\xi)}-\mathbf{r}_{(i,\xi')}$, and $\mu_{0}$ is the vacuum
permeability.

The defect electron spin couples to the nuclear spins through the hyperfine interaction of the form
\begin{equation}
H_{{\rm int}}=\sum_{(i,\xi)}\mathbf{S}\cdot\mathbb{A}_{(i,\xi)}\cdot\mathbf{I}^{(i,\xi)},
\end{equation}
with the coupling tensor
\begin{equation}
\mathbb{A}_{(i,\xi)}=\frac{\mu_{0}}{4\pi}\frac{\gamma_{e}\gamma_{\xi}}{r_{(i,\xi)}^{3}}\left[1-\frac{3\mathbf{r}_{(i,\xi)}\mathbf{r}_{(i,\xi)}}{r_{(i,\xi)}^{2}}\right].
\label{Eq:hf}
\end{equation}
Since the electron wave function of $\rm{T_{V2}}$ centers is quite localized~\cite{wave_function} (similar to the NV center case),
in Eq.~(\ref{Eq:hf}) we assume the hyperfine coupling being in dipolar form.
For nuclear spin bath of natural abundance, the typical hyperfine strength between the electron and the nuclear spin is $\lesssim 100~{\rm kHz}$
(corresponding to distance $r_{(i,\xi)} \gtrsim 5~\AA$). Since the hyperfine coupling is much smaller
than the ZFS, as long as the system is far from the level-crossing point, we can neglect the electron spin 
flipping process (i.e. the $S_x$ and $S_y$ terms) and, consequently, $S_z$ is a good quantum number taking the values $S_z=-3/2, \dots, 3/2$.
With this pure dephasing approximation, the hyperfine interaction  $H_{{\rm int}}$ is expanded as
\begin{eqnarray}
H_{{\rm int}}&\approx&\sum_{m=-3/2}^{3/2}\left|m\right\rangle \left\langle m\right|\otimes b_{m},\\
b_{m}&=&m \sum_{(i,\xi)} \hat{\mathbf{z}}\cdot\mathbb{A}_{(i,\xi)}\cdot\mathbf{I}^{(i,\xi)},
\end{eqnarray}
where $\left|m\right\rangle $ is the is the eigenstate of $H_{\rm{es}}$
with eigenvalue $\omega_{m}=m^{2}\Delta-m\gamma_{e}B$, and $\hat{\mathbf{z}}$ is the unit vector of the $z$ direction.

As the spin component $S_{z}$ is conserved, the population of each electron spin
states remains unchanged during the evolution. Now, we study the electron spin coherence defined as
\begin{equation}
L(t)=\frac{{\rm Tr}[\rho(t)S_{+}]}{{\rm Tr}[\rho(0)S_{+}]},
\end{equation}
where $S_{+}=S_{x}+iS_{y}$, and $\rho$ is density matrix of the
total system. The system is initially prepared in a product state
$\rho(0)=\rho_{\rm bath}\otimes\left|\psi_{e}(0)\right\rangle \left\langle \psi_{e}(0)\right|$,
where $\rho_{\rm bath}=\mathbb{I}^{N}/2^{N}$ is the density matrix of the bath spins with $2\times2$ identity
matrix $\mathbb{I}$, total bath spin number $N=N_{\rm C}+N_{\rm Si}$, and the electron spin initial state $\left|\psi_{e}(0)\right\rangle =(\left| m \right\rangle +\left| n\right\rangle)/\sqrt{2}$.
Coherence of different electron spin superposition states may have different decay time due to the back-action of electron spin to the bath spins, but  the overall decoherence time, the main concern in this paper, does not change significantly (in the same order of magnitude).
The back-action effect has been studied in NV center system both theoretically~\cite{PRL11_NZhao} and experimentally~\cite{NatComm11_JFDu}.
Here we chose the magnetic quantum number $m=3/2$ and $n=1/2$, as demonstrated in recent experiment~\cite{arxiv14_Wrachtrup}, 
to discuss the heteronuclear spin decoherence physics in SiC spin bath.
In the following,  we employ the cluster-correlation expansion (CCE) method\cite{PRB08_WenYang,PRB09_WenYang},
which is well-examined in similar systems such as NV centers in diamond~\cite{PRB12_NZhao} and
phosphorus donors in silicon (Si:P)~\cite{PRB07_Das Sarma}, to handle the $\rm V_{Si}$ decoherence problem in SiC nuclear spin bath.

\subsection{Heteronuclear spin pair dynamics\label{sub:Heteronuclear}}

\begin{figure}
\includegraphics[width=8cm]{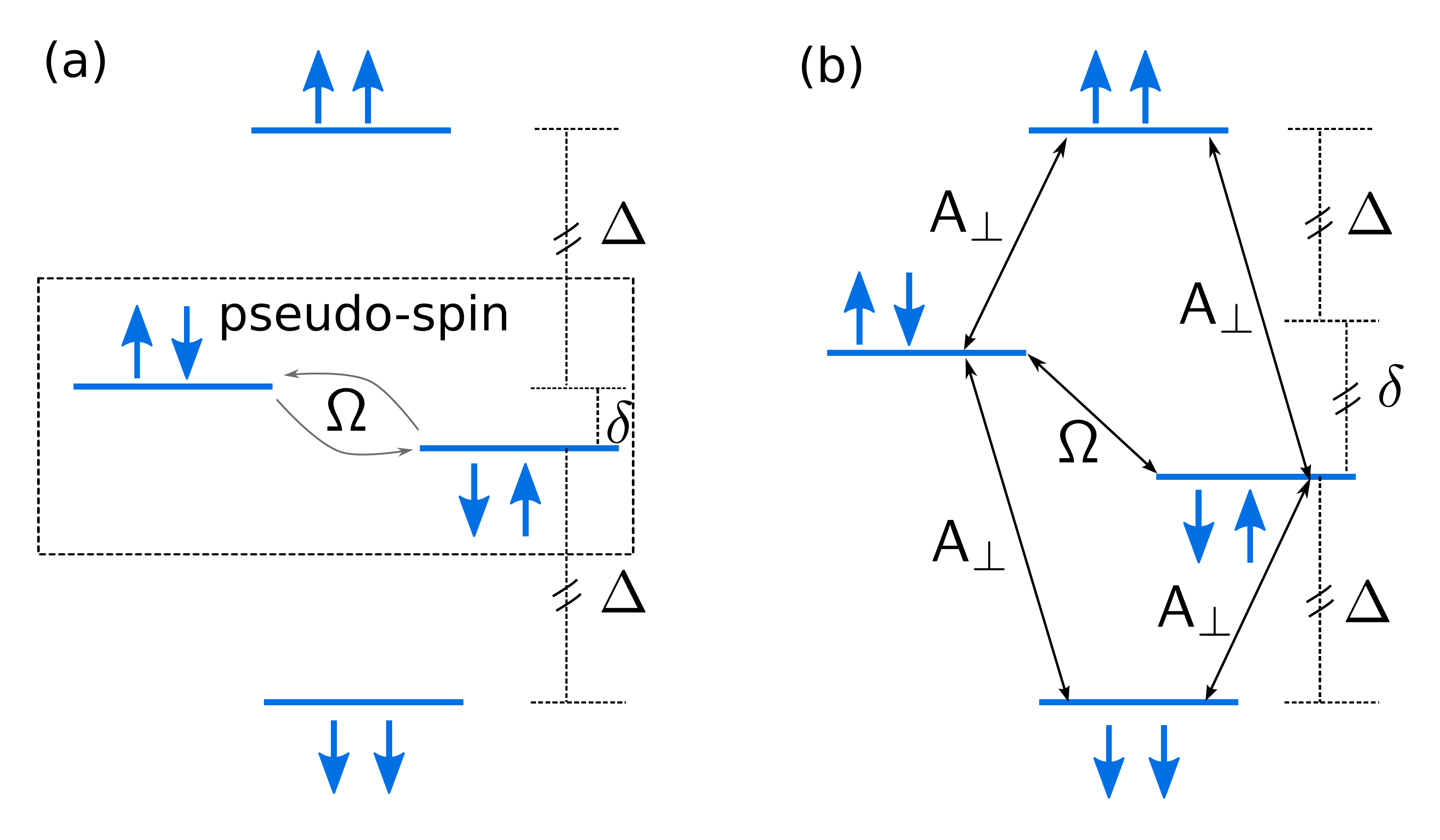}

\caption{\label{fig:pairlevel}(Color online) (a) Pseudo-spin model of homonuclear spin pair in strong magnetic field.
The polarized states are well-decoupled due to the large Zeeman splitting~$\Delta$. 
The unpolarized states form a pseudo-spin with frequency splitting~$\delta$
and transition rate~$\Omega$. 
(b) Energy levels of heteronuclear spin pair in the strong magnetic field. Both level splittings $\Delta$ and $\delta$ are proportional to magnetic field strength.
In additional to the secular flip-flop between the unpolarized states, the non-secular single spin flipping, induced by the vertical component of the hyperfine field~$A_{\perp}$, can also happen with comparable probability (see text).
However, all the spin transitions are significantly suppressed in strong magnetic fields. }
\end{figure}

Previous studies~\cite{PRB06_WangYao,PRL07_WangYao,NJP07_Liu,PRB03_Das Sarma,PRB05_Das Sarma,PRL07_Das Sarma}
showed that nuclear spin pair flip-flop is one of the
main decoherence mechanisms of an electron spin in a nuclear spins bath, particularly in strong
magnetic fields. For a given electron spin state $\left|m\right\rangle $, the Hamiltonian
of a nuclear spin pair is
\begin{equation}
H_{i\xi,j\xi'}^{(m)} =\mathbf{b}_{(i,\xi)}^{(m)}\cdot\mathbf{I}^{(i,\xi)}+\mathbf{b}_{(j,\xi')}^{(m)}\cdot\mathbf{I}^{(j,\xi')}+\mathbf{I}^{(i,\xi)}\cdot\mathbb{D}_{i\xi,j\xi'}\cdot\mathbf{I}^{(j,\xi')},
\end{equation}
where $\mathbf{b}_{(i,\xi)}^{(m)}=\mathbf{A}_{(i,\xi)}^{(m)}-\gamma_{\xi}B\hat{\mathbf{z}}$
is the effective magnetic field experienced by the $(i,\xi)$th nuclear
spin with the hyperfine field $\mathbf{A}_{(i,\xi)}^{(m)}\equiv m\hat{\mathbf{z}}\cdot\mathbb{A}_{(i,\xi)}$.
In the SiC nuclear spin bath, nuclear spin pair dynamics is different for the heteronuclear spin pairs 
(i.e., $\xi\neq\xi^{\prime}$ for $^{29}{\rm Si}$-$^{13}{\rm C}$ pairs)
to the homonuclear spin pairs (i.e., $\xi=\xi^{\prime}$ for $^{13}{\rm C}$-$^{13}{\rm C}$ and $^{29}{\rm Si}$-$^{29}{\rm Si}$ pairs).

\begin{figure}
\includegraphics[width=8cm]{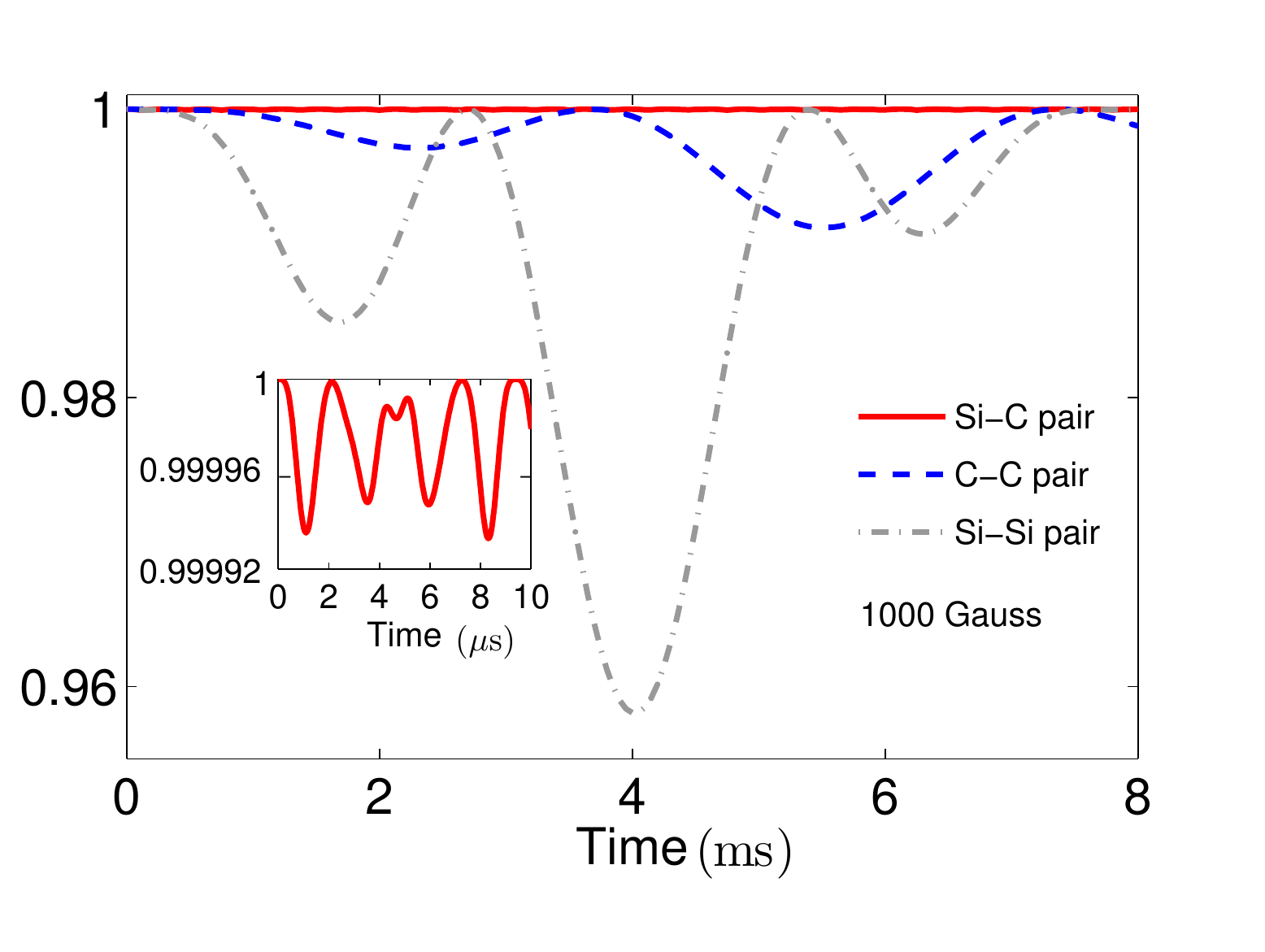}

\caption{\label{fig:singlepair}(Color online) The contributions of single nuclear spin pairs
to the electron spin coherence (Hahn echo). The red solid line, blue dashed line,
 and gray dashed-dotted line correspond to $^{29}\rm Si$-$^{13}\rm C$ pair, $^{13}\rm C$-$^{13}\rm C$ pair,
 and $^{29}\rm Si$-$^{29}\rm Si$ pair, in turn. The magnetic field is $B=1000\rm{Gauss}$.
 The inset shows a close-up of the contribution of $^{29}\rm Si$-$^{13}\rm C$ heteronuclear spin pair.
 All these pairs have similar inter-nuclei separations and distances to the defect center. }
\end{figure}

The contribution of homonuclear spin pairs to the central spin decoherence is well-studied in various systems like NV center and Si:P.
In strong magnetic fields, the dynamics of homonuclear spin pair can be described by a pseudo-spin model~\cite{PRB12_NZhao}.
The energy levels of nuclear spin pair in strong field are shown in Fig.~\ref{fig:pairlevel} (a).
The two polarized states (i.e., $\left|\uparrow\uparrow\right\rangle $
and $\left|\downarrow\downarrow\right\rangle $) are frozen by the large Zeeman energy~$\Delta$, and do not contribute to decoherence.
The two unpolarized states, $\left|\uparrow\downarrow\right\rangle $
and $\left|\downarrow\uparrow\right\rangle $, form a pseudo-spin (a two-level system) with a frequency splitting
$\delta$ and a transition rate $\Omega$.
For homonuclear spin pairs, as in the diamond spin bath of NV centers, the frequency 
splitting $\delta$ comes from the hyperfine field difference between the two nuclei (typically $\gtrsim {\rm kHz}$ for typical pairs with separation of several angstroms),
and the transition rate $\Omega$ is determined by 
the secular part of their dipole-dipole interaction (typically in the order of $\lesssim 10^2\rm{Hz}$).
The weak dipolar interaction between the two nuclei causes the spin flip-flop with a period $\sim{\rm ms}$, and results in the electron
spin decoherence in the order of $ (\Omega/\Delta)^2\sim 10^{-2}$ (see Fig.~\ref{fig:singlepair}).
A large number of such homonuclear spin pairs (about~$10^3$ pairs within
a large enough cut-off distance $R_{\rm c}=4~\text{nm}$) around the central electron spin contribute to the Hahn echo decay
on a timescale around $\sim\rm{ms}$.

In general, the dynamics of heteronuclear spin pair cannot be well-described by a pseudo-spin model as in the homonuclear spin case.
For heteronuclear spin pairs, the splitting $\delta$ also consists of the Zeeman frequency
difference due to the different gyromagnetic ratios ($\vert \gamma_{\rm C}\vert - \vert\gamma_{\rm Si}\vert= 2\pi \times 0.2~{\rm kHz/Gauss}$). 
The Zeeman frequency difference is usually much larger than the hyperfine field difference in the strong fields (e.g. $B>300~{\rm Gauss}$).
In this case, both level splittings $\Delta$ (between polarized state and unpolarized state) and $\delta$ (between the two unpolarized states)  are proportional to magnetic field strength.
The non-secular transition probability (characterized by the ratio $A_{\perp}/\Delta$ with $A_{\perp}$ being the component of the hyperfine field difference perpendicular to the magnetic field) could be larger than the secular transition probability (characterized by the ratio $\Omega/\delta$).
Thus, the levels of heteronuclear spin pairs [see Fig.~\ref{fig:pairlevel} (b)] cannot be simplified to a pseudo-spin model by neglecting the polarized spin states.
On the other hand, in strong fields, both secular and non-secular spin transitions are significantly suppressed 
(i.e. $A_{\perp}/\Delta\ll 1$ and $\Omega/\delta\ll 1$), and the contribution to electron spin decoherence of heteronuclear 
spin pairs is negligibly small on a time scale of $\sim{\rm ms}$ (much smaller than the homonuclear spin cases, see Fig.~\ref{fig:singlepair}).
With the heteronuclear spin pair dynamics suppressed, the electron spin decoherence is, indeed,
caused by two independent baths ($^{13}{\rm C}$ and $^{29}{\rm Si}$ spin baths).
The effective nuclear spin concentration of these two baths are reduced by a factor of $\sim 2$, which
greatly prolongs the decoherence time of defect center spins as shown in the next section.


In the weak external magnetic region ($B\ll10~{\rm Gauss}$), the heteronuclear
and homonuclear spin pairs have similar contribution to the decoherence. In this region,
the Zeeman splitting of the nuclear spins $\gamma_{\xi}B$ is much smaller than
the hyperfine interaction strength.
The requirement of Zeenman energy conservation does not hold, and the non-secular spin flipping processes (e.g., transition between $\vert\uparrow \uparrow\rangle$ and $\vert\uparrow \downarrow\rangle$ states) become important.
As the difference in the gyromagnetic ratio plays little role in the spin pair dynamics, both the hetero- and homonuclear spin pairs
contribute similarly to the electron spin decoherence.
A detailed study the spin pair dynamics in weak magnetic fields can be found in Ref.~\onlinecite{PRB12_NZhao}.

\section{results and discussion\label{sec:results}}

\subsection{Free-induced decay}

\begin{figure}
\includegraphics[width=8cm]{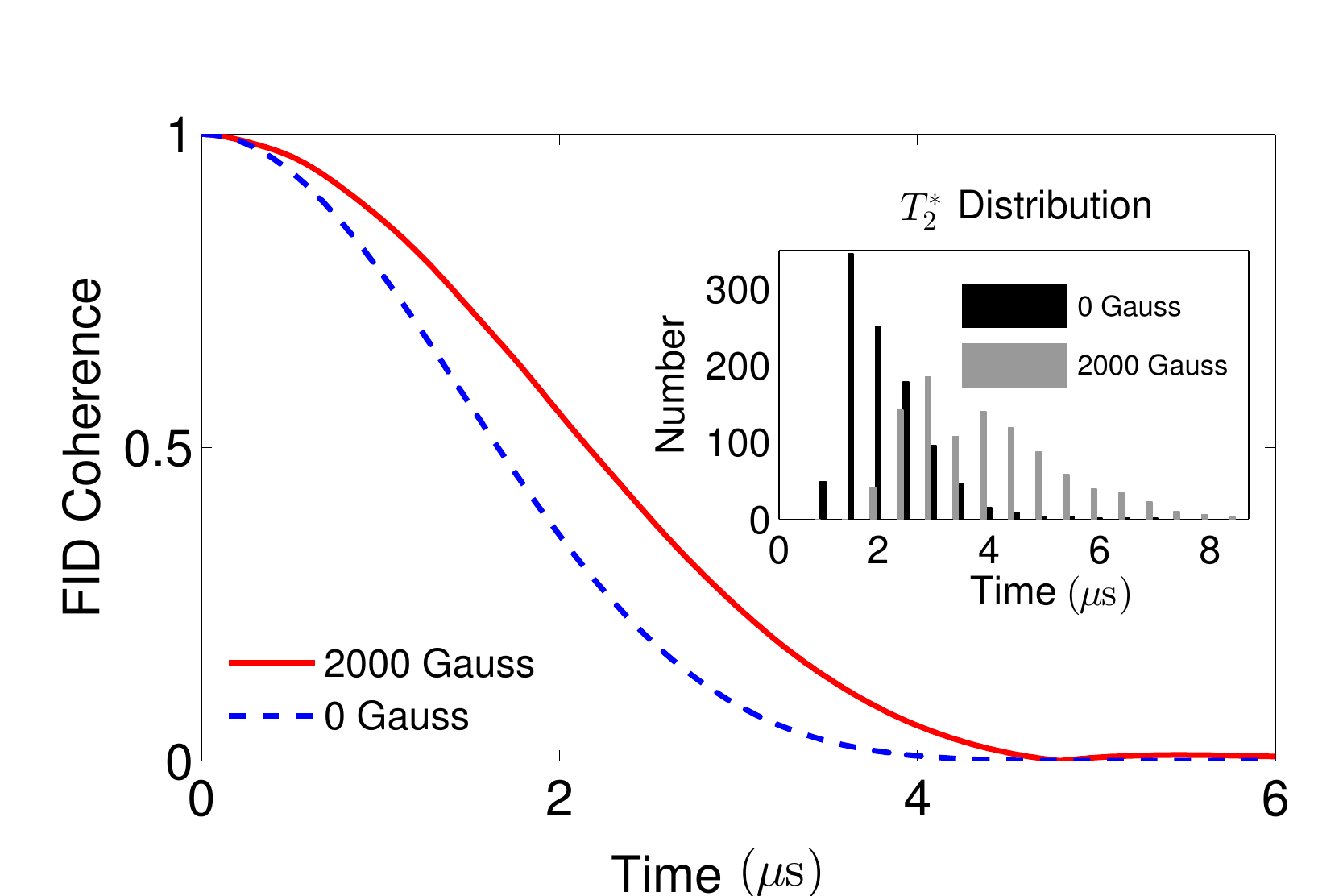}

\caption{\label{fig:FID}(Color online) The typical FID of $\rm{T_{V2}}$ center
in the cases of zero field (dashed blue line) and strong field (solid red line)
are presented. The inset gives the histogram of FID coherence time
$T_{2}^{*}$ distribution of 1000 randomly generated bath configurations
under zero and strong fields.}
\end{figure}

In this section, we show the numerical results of the electron spin coherence of ${\rm T_{V2}}$ centers.
We start from the free-induction decay (FID) of the central spin.
Identical to the NV center case, in both weak field and strong field cases, the FID
is of Gaussian shape as shown in Fig.~\ref{fig:FID}. The inset of
Fig.~\ref{fig:FID} shows a histogram of FID coherence times  $T_{2}^{*}$ for 1000 randomly generated
bath spin configurations under zero and strong magnetic fields.
The mean decoherence times over different configurations under zero
and strong field are $T_{2}^{*}\approx1.9~{\rm \mu s}$ and $T_{2}^{*}\approx3.3~{\rm \mu s}$,
respectively, which are both comparable to the corresponding decoherence
times $T_{2}^{*}\approx2.1~{\rm \mu s}$ and $T_{2}^{*}\approx3.6~{\rm \mu s}$ of
the NV centers given in Ref.~\cite{PRB12_NZhao} and Ref.~\cite{NJP12_Maze}.

The electron spin FID coherence time is inversely proportional to the concentration
of nuclear spins~\cite{PRB05_Das Sarma,PRL07_Das Sarma}. 
Notice that the natural abundance of $^{29}\rm{Si}$ is about 4 times larger than that of $^{13}\rm{C}$.
The reasons for the $T_2^*$  time of $\rm T_{V2}$ centers is not reduced significantly are
(i) the~${\rm C-Si}$ bond length $1.89~\AA$ in SiC is lager than the ${\rm C-C}$ bond length $1.54~\AA$ in diamond, which implies the
volume density of nuclear spins is reduced by a factor of~$(1.89/1.54)^3 = 1.8$;
(ii) about $80\%$ of the nuclear spins in the bath are $^{29}{\rm Si}$, which have smaller gyromagnetic ratios than
$^{13}\rm{C}$ ($|\gamma_{\rm C}/\gamma_{\rm Si}| \approx 1.3$) and, as a result, produce weaker hyperfine fluctuations.
These two factors compensate the larger natural abundance of the $^{29}{\rm Si}$, and results in similar $T_2^*$ times  of
$\rm{T}_{\rm{V2}}$ and NV centers in diamond.

\subsection{Hahn echo}

\begin{figure*}
\includegraphics[width=15cm]{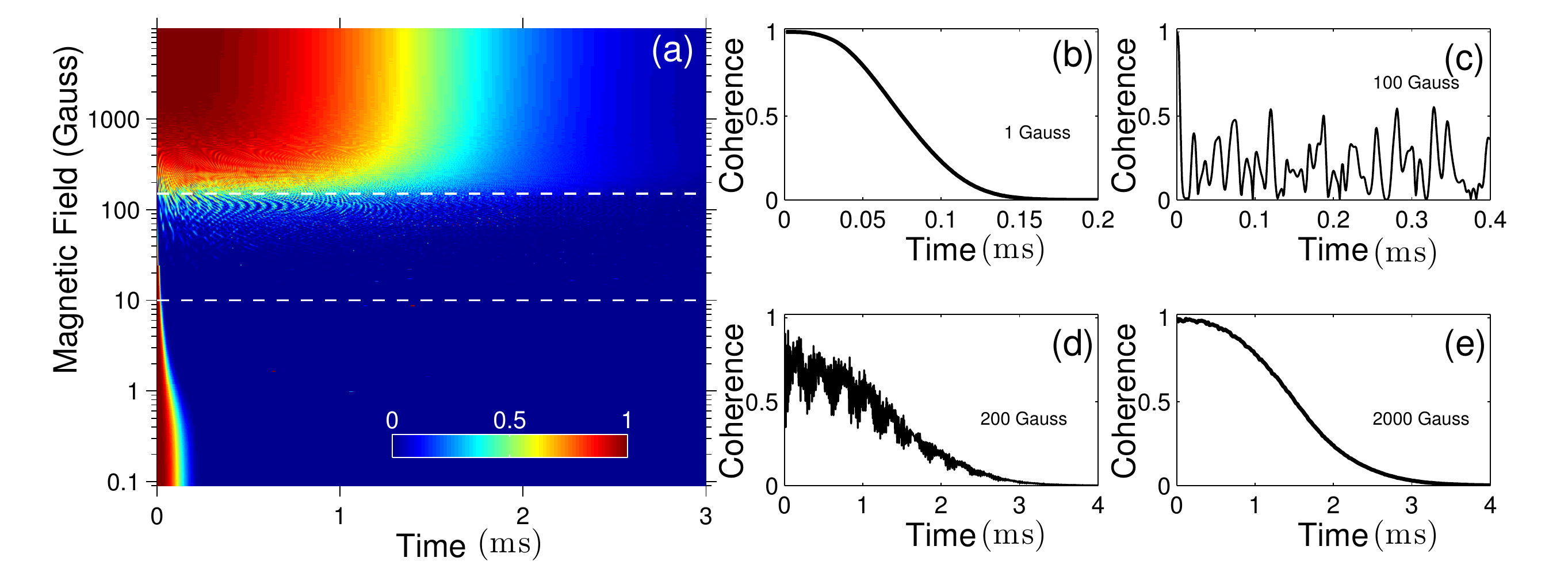}\caption{\label{fig:Hahnecho2D}(Color online)
(a) Hahn echo of $\rm{T_{V2}}$ center spin
coherence between $\left|1/2\right\rangle $ and $\left|3/2\right\rangle $ states
as a function of total evolution time $t$ and magnetic field $B$.
The strong, medium, and weak magnetic field regimes, in which the
$\rm{T_{V2}}$ center spin has different decoherence behavior, are separated by
the horizontal dashed lines. (b-e)  the typical Hahn echo coherence
behavior for $1~\rm{Gauss}$, $100~\rm{Gauss}$, $200~\rm{Gauss}$, and $2000~\rm{Gauss}$ magnetic fields, in turn.}
\end{figure*}

Figure~\ref{fig:Hahnecho2D} (a) shows the electron spin coherence under spin echo (Hahn echo) control in different magnetic fields.
Similar to the NV center case, the electron spin decoherence of $\rm T_{V2}$ centers is qualitatively different in different magnetic field regimes, namely, the weak, medium, and strong regimes.
In the weak magnetic field regime ($B \ll 10~{\rm Gauss}$),
the coherence decays monotonically within $\sim 200~{\rm \mu s}$ [see Fig.\ref{fig:Hahnecho2D} (b)].
The decay time is comparable to that of NV centers in the same weak fields,
which once again shows that the effective nuclear spin concentration of SiC bath in weak fields is similar to the diamond spin bath, 
due to the unit volume expansion and the smaller gyromagnetic ratio of $^{29}{\rm Si}$.

In medium magnetic fields ($10~{\rm Gauss} \lesssim B\lesssim 150~{\rm Gauss}$),
the coherence collapses on a short time scale ($\sim 10~{\rm \mu s}$) and partially revives at later time [see Fig.~\ref{fig:Hahnecho2D} (c)].
The revival pattern is quite irregular and is sensitive to the random configuration of nuclear spins close to the $\rm T_{V2}$ center.
This is different from the situations of NV center in diamond~\cite{Science06_Lukin} and the divacancy center in SiC~\cite{Nature_ Awschalom}, where spin echo coherence collapses and revives either
periodically (for NV center), or with a regular beating pattern determined by the two Larmor frequencies of $^{13}\rm{C}$ and $^{29}\rm{Si}$ (for divacancy).
In the $\rm T_{V2}$ center case,  the irregular modulation comes form the fact that the electron spin is spin-$3/2$ (half integer).
The all 4 states (i.e., $\vert m \rangle$ with $m=3/2, 1/2, -1/2$ and $-3/2$) of $\rm T_{V2}$ electron spin have 
different {\it non-zero} hyperfine couplings to each nuclear spins,
which modifies their precession frequencies.
The frequency modifications depend on the particular positions of each nuclear spins.
Consequently, there does not exist a common precession frequency for all the bath spins, resulting in the irregular revival patten.
In contrast, in the NV center case or divacancy case (both are integer spin-$1$),
for electron spin in $\vert m=0\rangle$ state, all nuclear spins have only one
(for NV center) or two (for divacancy in SiC) common precession frequencies (in the absence of hyperfine coupling),
which is essential for periodic or regular coherence revivals.

In the strong-field region ($B\gg150{\rm Gauss}$), the collapse
and revival effects are greatly suppressed [see Fig.~\ref{fig:Hahnecho2D} (d)] and finally vanish in the strong external field limit {[see Fig.~\ref{fig:Hahnecho2D} (e)].
The coherence monotonically decays as in the weak field
regime, but with a longer time scale more than $\sim1~\rm{ms}$, which is almost
twice as long as the typical coherence time of NV center in diamond spin bath.

\begin{figure}
\includegraphics[width=8.5cm]{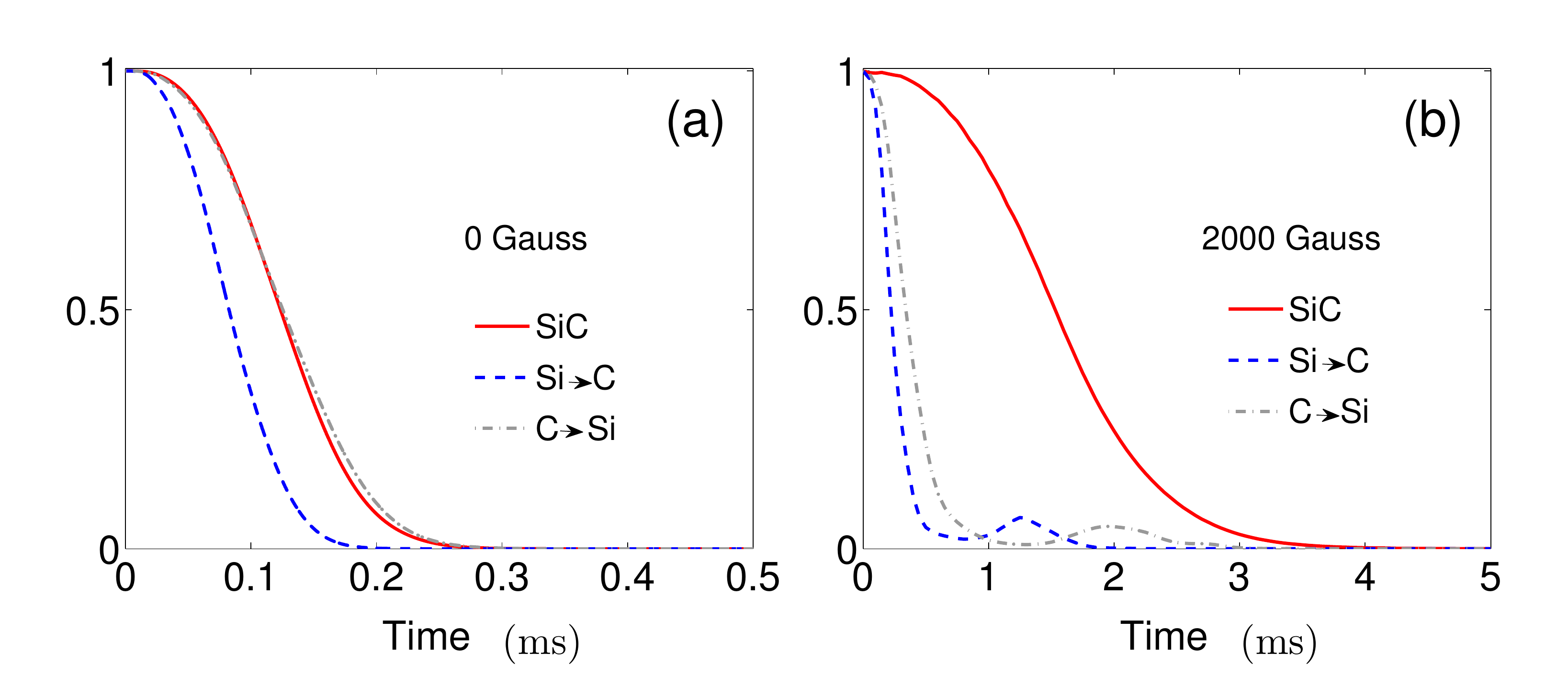}

\caption{\label{fig:pairmenchanism}(Color online) (a) The influence of heteronuclear spin pairs to the Hahn echo in zero field.
The red solid line is the electron spin coherence in real SiC nuclear spin bath.
The blue dashed line is calculated with all $^{29}{\rm Si}$ spins replaced by $^{13}{\rm C}$ spins.
The gray dashed-dotted line is calculated with all the $^{13}{\rm C}$ spins replaced by $^{29}{\rm Si}$ spins.
(b) The same as (a), but in strong magnetic field.}
\end{figure}

The reason for the longer coherence time in strong magnetic field is the suppression of heteronuclear spin pair flip-flop process.
As we discussed in Sect.~\ref{sub:Heteronuclear}, in the strong magnetic field regime, $^{13}{\rm C}$ and $^{29}{\rm Si}$ nuclear spins form two independent baths.
The spin concentrations of the two independent baths are twice smaller than if all the nuclear spins were of the same isotope.
To further prove this bath dilution mechanism, we calculate the electron spin decoherence with all $^{29}{\rm Si}$ spins replaced by $^{13}{\rm C}$ spin or vice versa ($^{13}{\rm C}$ replaced by $^{29}{\rm Si}$) while keeping all the other conditions (e.g., nuclear spin positions, magnetic field strength, etc.) unchanged.
Figure~\ref{fig:pairmenchanism} (b) shows that with either replacements, the coherence time will be significantly reduced due to the opening of the decoherence channel between two originally independent baths.
The coherence time does not change significantly if we do the same isotope replacement in the weak field regime [see Fig.~\ref{fig:pairmenchanism} (a)] because, as we analyzed above, hetero- and homonuclear spin pairs have similar contributions to decoherence.
With this we conclude that the different behavior of heteronuclear spin pairs in different magnetic fields is the key point for understanding the coherence time of defect centers in SiC.

\section{conclusion\label{sec:conclusion}}

We investigate the electron spin coherence time of single defects (e.g., $\rm{T_{V2}}$ centers) in SiC nuclear spin bath in different magnetic fields.
Our results show that the defect centers in SiC can have longer coherence  time than the NV center in diamond even though the natural abundance of $^{29}{\rm Si}$ is higher than $^{13}{\rm C}$.
Through numerical calculations based on a microscopic model, we analyze the decoherence mechanisms,
and find the longer coherence time is the consequence of the suppression of the heteronuclear spin pair flip-flop process.
Our work confirms that electron spins of defect center in SiC are excellent candidates for quantum
information processing and future spin-based quantum devices.

\section*{Acknowledgement}
L.P.Y. and C.B. equally contributed to this work. This work is supported by NKBRP (973 Program) 2014CB848700, and NSFC No. 11374032 and No. 11121403.  
J. W. acknowledges  Max Planck Society, EPR priority program of the DFG, EU via SQUTEC.

\end{document}